\documentclass{article}
\usepackage{courier}
\usepackage{url}
\usepackage{longtable}
\usepackage{multirow}
\usepackage{float}
\usepackage{subfloat}
\usepackage{lscape,epsfig}
\usepackage{graphicx}
\usepackage{amssymb}
\usepackage{amsmath}
\usepackage{natbib}
\textheight=22cm \DeclareSymbolFont{ppa}{OT1}{ppl}{m}{it}
\DeclareMathSymbol{\vv}{\mathalpha}{ppa}{'166}

minus 2.3mu \thickmuskip = 2.6mu plus 2mu minus 2.6mu

\let\svthefootnote\thefootnote

\begin{document}

\newcommand{\dd}{\,{\rm d}}
\newcommand{\ie}{{\it i.e.},\,}
\newcommand{\etal}{{\it $et$ $al$.\ }}
\newcommand{\eg}{{\it e.g.},\,}
\newcommand{\cf}{{\it cf.\ }}
\newcommand{\vs}{{\it vs.\ }}
\newcommand{\zdot}{\makebox[0pt][l]{.}}
\newcommand{\up}[1]{\ifmmode^{\rm #1}\else$^{\rm #1}$\fi}
\newcommand{\dn}[1]{\ifmmode_{\rm #1}\else$_{\rm #1}$\fi}
\newcommand{\upd}{\up{d}}
\newcommand{\uph}{\up{h}}
\newcommand{\upm}{\up{m}}
\newcommand{\ups}{\up{s}}
\newcommand{\arcd}{\ifmmode^{\circ}\else$^{\circ}$\fi}
\newcommand{\arcm}{\ifmmode{'}\else$'$\fi}
\newcommand{\arcs}{\ifmmode{''}\else$''$\fi}
\newcommand{\MS}{{\rm M}\ifmmode_{\odot}\else$_{\odot}$\fi}
\newcommand{\RS}{{\rm R}\ifmmode_{\odot}\else$_{\odot}$\fi}
\newcommand{\LS}{{\rm L}\ifmmode_{\odot}\else$_{\odot}$\fi}

\newcommand{\Abstract}[2]{{\footnotesize\begin{center}ABSTRACT\end{center}
\vspace{1mm}\par#1\par \noindent {~}{\it #2}}}

\newcommand{\TabCap}[2]{\begin{center}\parbox[t]{#1}{\begin{center}
  \small {\spaceskip 2pt plus 1pt minus 1pt T a b l e}
  \refstepcounter{table}\thetable \\[2mm]
  \footnotesize #2 \end{center}}\end{center}}

\newcommand{\TableSep}[2]{\begin{table}[p]\vspace{#1}
\TabCap{#2}\end{table}}

\newcommand{\FigCap}[1]{\footnotesize\par\noindent Fig.\  %
  \refstepcounter{figure}\thefigure. #1\par}

\newcommand{\TableFont}{\footnotesize}
\newcommand{\TableFontIt}{\ttit}
\newcommand{\SetTableFont}[1]{\renewcommand{\TableFont}{#1}}
\newcommand{\MakeTable}[4]{\begin{table}[htb]\TabCap{#2}{#3}
  \begin{center} \TableFont \begin{tabular}{#1} #4
  \end{tabular}\end{center}\end{table}}

\newcommand{\MakeTableSep}[4]{\begin{table}[p]\TabCap{#2}{#3}
  \begin{center} \TableFont \begin{tabular}{#1} #4
  \end{tabular}\end{center}\end{table}}

\newenvironment{references}%
{ \footnotesize \frenchspacing
\renewcommand{\thesection}{}
\renewcommand{\in}{{\rm in }}
\renewcommand{\AA}{Astron.\ Astrophys.}
\newcommand{\AAS}{Astron.~Astrophys.~Suppl.~Ser.}
\newcommand{\ApJ}{Astrophys.\ J.}
\newcommand{\ApJS}{Astrophys.\ J.~Suppl.~Ser.}
\newcommand{\ApJL}{Astrophys.\ J.~Letters}
\newcommand{\AJ}{Astron.\ J.}
\newcommand{\IBVS}{IBVS}
\newcommand{\PASP}{P.A.S.P.}
\newcommand{\Acta}{Acta Astron.}
\newcommand{\MNRAS}{MNRAS}
\renewcommand{\and}{{\rm and }}
\section{{\rm REFERENCES}}
\sloppy \hyphenpenalty10000
\begin{list}{}{\leftmargin1cm\listparindent-1cm
\itemindent\listparindent\parsep0pt\itemsep0pt}}%
{\end{list}\vspace{2mm}}

\def\TYLDA{~}
\newlength{\DW}
\settowidth{\DW}{0}
\newcommand{\dw}{\hspace{\DW}}

\newcommand{\refitem}[5]{\item[]{#1} #2%
\def\REFARG{#3}\ifx\REFARG\TYLDA\else, {\it#3}\fi
\def\REFARG{#4}\ifx\REFARG\TYLDA\else, {\bf#4}\fi
\def\REFARG{#5}\ifx\REFARG\TYLDA\else, {#5}\fi.}

\newcommand{\Section}[1]{\section{#1}}
\newcommand{\Subsection}[1]{\subsection{#1}}
\newcommand{\Acknow}[1]{\par\vspace{5mm}{\bf Acknowledgements.} #1}
\pagestyle{myheadings}

\newfont{\bb}{ptmbi8t at 12pt}
\newcommand{\xrule}{\rule{0pt}{2.5ex}}
\newcommand{\xxrule}{\rule[-1.8ex]{0pt}{4.5ex}}

\begin{center}
{\Large\bf
 The Cluster AgeS Experiment (CASE).\dag  \\
 Detecting Photometric Variability with the Friends of Friends Algorithm}{\LARGE$^\ast$}
 \vskip1cm
  {\large
      ~~M.~~R~o~z~y~c~z~k~a$^1$,
      ~~W.~~N~a~r~l~o~c~h$^1$,
      ~~P.~~P~i~e~t~r~u~k~o~w~i~c~z$^2$
      ~~I.~B.~~T~h~o~m~p~s~o~n$^3$~~and
      ~~W.~~P~y~c~h$^1$
   }
  \vskip3mm
{ $^1$Nicolaus Copernicus Astronomical CENter of the Polish Academy of Sciences, 
  ul. Bartycka 18, 00--716 Warsaw, Poland\\
     e--mail: (mnr, pych, wnarloch)@camk.edu.pl\\
  $^2$Warsaw University Observatory, Al. Ujazdowskie 4, 00-478 Warsaw, Poland\\
     e--mail: pietruk@astrouw.edu.pl\\
  $^3$The Observatories of the Carnegie Institution for Science, 813 Santa Barbara
      Street, Pasadena, CA 91101, USA\\
     e--mail: ian@obs.carnegiescience.edu\\
}
\end{center}

\vspace*{7pt}
\Abstract 
{We adapt the friends of friends algorithm to the analysis of light curves, and show that it can be 
succesfully applied to searches for transient phenomena in large photometric databases. As a test case 
we search OGLE-III light curves for known dwarf novae. A single combination of control parameters 
allows to narrow the search to 1\% of the data while reaching a $\sim$90\% detection efficiency. A 
search involving $\sim$2\% of the data and three combinations of control parameters can be significantly 
more effective - in our case a 100\% efficiency is reached.

The method can also quite efficiently detect semi-regular or strictly periodic variability. We report 28 
new variables found in the field of the globular cluster M22, which was examined earlier with the help of 
periodicity-searching algorithms. 
}
{Methods: data analysis -- Stars: variables -- globular clusters: individual (M22) 
-- Galaxy: disk – binaries: close – novae, cataclysmic variables
}

\let\thefootnote\relax\footnotetext{\dag CASE was initiated and for long time led
by our friend and tutor Janusz Kaluzny, who prematurely passed away in March 2015.}
\let\thefootnote\relax\footnotetext
{$^{\mathrm{\ast}}$Based on data obtained with Swope and Warsaw telescopes at  
Las Campanas Observatory.}
\let\thefootnote\svthefootnote

\Section{Introduction} 
\label{sec:intro}
The friends-of-friends method (FoF) was introduced in cosmology by Turner \& Gott (1976) 
as a tool to identify clusters of galaxies. In FoF, a cluster is defined using the 
clustering length $CL$ such that each member galaxy has at least one neighbor at a distance 
$d\le CL$. In other words, for each galaxy all neighbors closer than the $CL$-threshold are 
members of the cluster which that galaxy belongs to. 

The basic algorithm of the method was published by Huchra~\& Geller (1982). Although 
several more sophisticated approaches to the clustering problem have been developed, 
FoF is still widely used for its simplicity and effectiveness. Its another advantage is
that it uniquely assigns galaxies to clusters, while making no assumptions on cluster 
properties. Obviously, for a given sample of galaxies both the number and the richness of 
FoF clusters depend on $CL$, so that a calibration is necessary. In fact, beginning with 
Huchra~\& Geller (1982), two clustering lengths have been used in most implementations:  
$CL_{\perp}=b_{\perp}*s(z)$ for distances in the plane of the sky, and $CL_{\parallel}=
b_{\parallel}*s(z)$ for distances along the line of sight, where $s(z)$ is the 
redshift-dependent mean three-dimensional separation between galaxies, and ($b_{\perp}$, 
$b_{\parallel}$) are scaling parameters to be calibrated. Surveys performed 
since 2010 and reviewed by Duarte \& Mamon (2014) yielded $b_{\perp}$ and $b_{\parallel}$ 
ranging from 0.06 to 0.11 and from 0.67 to 1.1, respectively, with $b_{\parallel}$  
consistently larger than $b_{\perp}$ to account for redshift space distortions (the so-called 
Fingers of God). In the same paper Duarte \& Mamon derived $b_{\perp} \simeq0.11$ and 
$b_{\parallel}\approx1.3$ using a mock galaxy catalog generated from cosmological 
simulations. 

The aim of the present contribution is to show that FoF can be successfully applied to the 
search for aperiodic variations in long series of photometric measurements. In the epoch
of massive photometry and robotic telescopes many algorithms capable of detecting this type 
of variability, with particular emphasis on sudden brightenings, have been developed. A
related review has recently been published by Sokolovsky et al. (2017), and there is no need 
to repeat their extensive discussion. We only mention that they test 18 statistical 
characteristics of brightness measurements, and compare their performance in identifying 
variable objects on several data sets with time-series photometry. FoF-based search for
variability can be a useful additional technique.

The implementation of the algorithm is outlined in Section \ref{sec:impl}. Control parameters of FoF
are calibrated in Section \ref{sec:calib}, and in Section \ref{sec:search} the method is 
applied to lightc urves collected within CASE in the field of the globular cluster M22. In 
Section~\ref{sec:ogle} we estimate the detection efficiency of FoF based on OGLE-III light curves 
surveyed by Mr\'oz et al. (2013, hereafter MP13), and Pietrukowicz et al. (2013). A brief summary of 
the paper is presented in Section~\ref{sec:sum}.
 
\section{Implementation of the algorithm}
\label{sec:impl}

\begin{figure}[H]
   \centerline{\includegraphics[width=0.95\textwidth,
               bb = 55 499 562 687, clip]{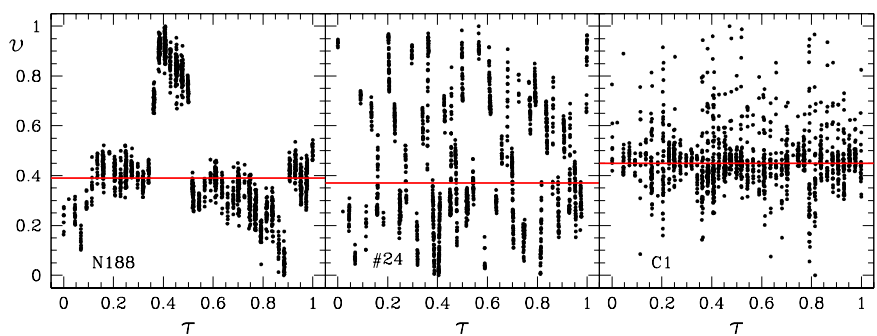}}
   \caption{Examples of normalized light curves (see text for explanations). Red lines mark median 
    values of $\vv$. Labels within the frames identify stars described in Section \ref{sec:calib}.
    \label{fig:norm}}
\end{figure}

Our original idea was to write a code capable of identifying irregularly``flashing'' objects,
e.g. dwarf novae (DN), and the mathematical problem we posed was the following: given a set 
$\cal S$ of time-sorted measurements $V_i(t_i)$ 
\begin{itemize}
\item divide it into clusters $G_k$ of points $p_{i} = (t_{i},V_{i})$ such that 
\begin{displaymath} 
p_{i} \in G_k \wedge d_{ij}\leq CL\ \Rightarrow\ p_j \in G_k,
\end{displaymath}
where $d_{ij}$ is the distance between $p_i$ and $p_j$, and $CL$ is the assumed 
clustering length, 
\item and identify the brightest, or ``peak'' cluster $G_p$ which most strongly deviates from the median 
$\mathrm{Md}(V)$ of all measurements in $\cal S$. 
\end{itemize}
To find the solution, a reasonable definition of distances is necessary. 
Our approach involves the following steps:
\begin{enumerate}
\item Compress the time axis by removing all Julian days without measurements.
Define the normalized time $\tau_i = (t_i-t_1)/(t_L-t_1)$, where $t_L$ 
and $t_1$ are Julian dates of the last and first measurement, respectively.
Remove points with $V>\mathrm{Md}(V)$ (this optional operation saves the CPU time, 
and biases the variability search towards brightenings which we are mainly interested in). 
Define the normalized magnitude $\vv_i=(\mathrm{Md}(V)-V_i)/(\mathrm{Md}(V)-V_{min})$, where $V_{min}$ 
is the magnitude of the brightest point in $\cal S$ (note that $\vv$ {\it increases} with increasing 
brightness).  All the points are now located within a unit square (see Fig. \ref{fig:norm}), and the 
distance $d_{ij}$ can be defined as the Cartesian distance 
between $(\tau_i,\vv_i)$ and $(\tau_j,\vv_j)$.
\item Compute all distances $d_{ij}$. Find the mean distance $\overline d$, and the clustering 
length $CL=b\overline d$, where $b$ is a control parameter corresponding to $b_\perp$ 
from Sect. \ref{sec:intro}. Draw a virtual circle of radius $CL$ around each $p_i$. 
Merge each set of overlapping circles into a cluster. 
\item Reject clusters comprising
fewer than $N_{\rm M}$ points, where $N_{\rm M}$ is another control parameter. 
For each remaining cluster $G_k$ find the minimum $\vv_k^{min}$, and the mean $\overline \vv_k$. 
Identify the peak cluster $G_p$ as the one for which $\vv_p^{min}=\max\limits_{k}(\vv_k^{min})$.
\end{enumerate}
Based on $\vv_p^{min}$ and the corresponding $\overline\vv_p$, the variability index $\nu$ can be 
calculated, which we define as
 \[
    \nu=
        \begin{cases}
          \vv_p^{min}\ \mathrm{if}\ \vv_p^{min}<0.4\\ 
         \overline\vv_p\ \mathrm{otherwise}.
        \end{cases}
  \]
Adopting such a definition, we have $1.0\ge\nu\ge0.0$, where larger values indicate the presence of at 
least one brightening episode, and smaller values suggest that the examined star does not vary 
in brightness. The cutoff at $\vv_p^{min}=0.4$ is introduced based on calibration results from 
the next Section, and its aim is to avoid assigning $\nu>0$ to stars with $\vv_p^{min}=0$. 

Note that the smaller $b$ is, the more densely populated areas of the $(\tau,\vv)$ 
plane are only sampled, and, for a nonvariable source, the smaller is the chance to generate 
spurious clusters with $\vv_k^{min}>0$ which would artificially increase the variability index. 
Obviously, if $b$ is too small for a given $N_{\rm M}$ then no clusters will be found. 
On the other hand, when $b$ is too large, real clusters with $\vv_k^{min}>0$ may be 
spuriously extended to include points with $\vv\approx 0$, causing the physical 
variability to be classified as a noise. In any case, the control parameters $b$ and $N_{\rm M}$ 
should be so chosen that $G_p$ contains sufficiently many points to represent a real physical 
increase in brightness. 

\Section{Calibrating FoF}
\label{sec:calib}

For the calibration, observations of the globular cluster M22 collected within the CASE project
(Kaluzny et al. 2005), and searched for periodic variability by Rozyczka et al. (2017; hereafter 
R17) were used. R17 examined light curves of almost 124,000 stars with $V \lesssim22$ mag, 
identifying over 350 periodic or likely periodic variables, and a few likely 
long--term variables. 238 of these were new detections.\footnote{Data for all the identified 
variables are available at http://case.camk.edu.pl} 

\begin{figure}[H]
   \centerline{\includegraphics[width=0.95\textwidth,
               bb = 34 299 563 696, clip]{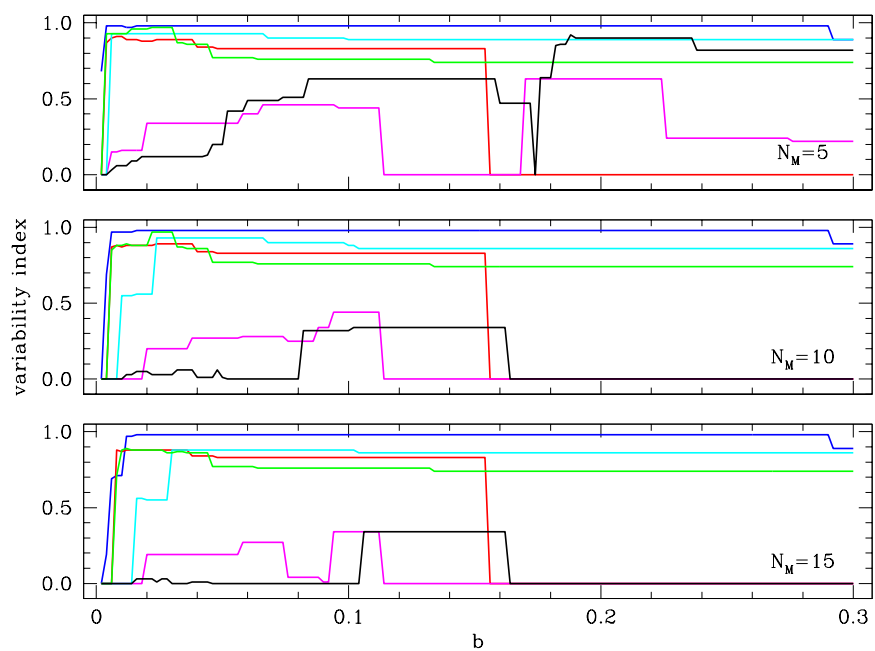}}
   \caption{ Variability index as a function of clustering length parameter $b$ and minimum
    size of the cluster $N_{\rm M}$ for stars U70, N188, N184, \#24, V136 and C1 from the M22 field 
    (red, green, blue, cyan, magenta and black curves, respectively). 
    \label{fig:scan}}
\end{figure}

Trial runs performed on various subsets of this data with several values of clustering parameter $b$ 
and minimum size of the cluster $N_{\rm M}$ yielded four new variables, which together with a BL 
Her - type pulsator described by R17 (hereafter star \#24), and a randomly picked constant member of 
M22 (hereafter star C1), provided 
input to the calibration procedure. The aim of the calibration was to find optimal values of
$b$ and $N_{\rm M}$, at which the physical variability is most easily detectable. The results are 
shown in Fig.~\ref{fig:scan}, where $\nu$ is plotted for all the six stars as a function of $b$ 
for $N_{\rm M}$ = 5, 10 and 15. 

\begin{figure}[H]
   \centerline{\includegraphics[width=0.95\textwidth,
               bb = 27 295 563 687, clip]{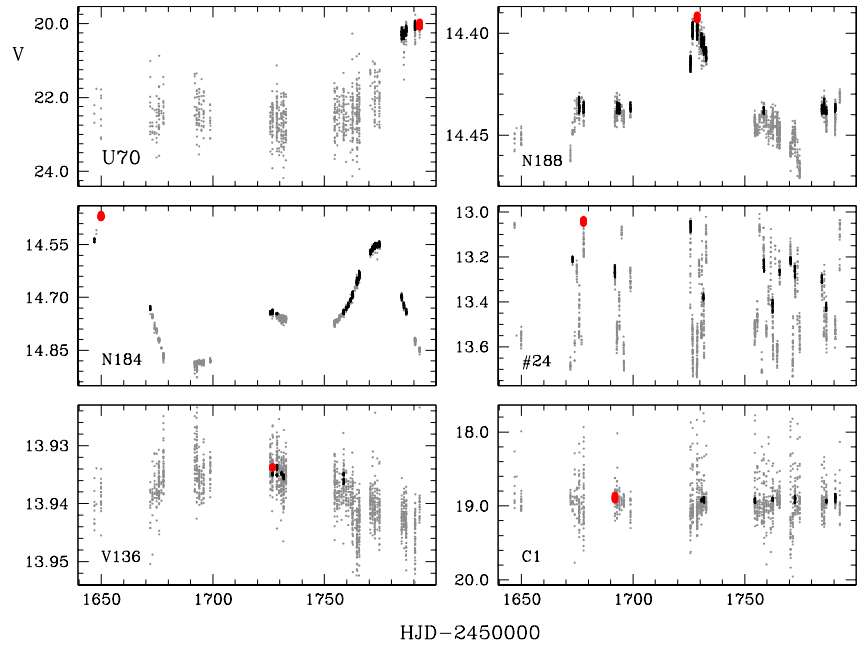}}
   \caption{Top left to bottom right: raw curves of stars U70, 
    N188, N184, \#24, V136 and C1. Grey: points located below the median 
    Md($V$) or belonging to clusters with less than $N_{\rm M}=10$ members. Black 
    or red: points belonging to clusters with at least $N_{\rm M}$ members. Red: the brightest 
    cluster, based on which the variability index is defined. Results obtained 
    for $b=0.03$.
    \label{fig:examples1003}}
\end{figure}

Stars U70 and N188 (red and green lines), which have evidently ``peaking'' light curves 
(see Fig. \ref{fig:examples1003}), nicely illustrate the dependence of $\nu$ on $b$ disussed in 
Sect. \ref{sec:impl}. As the clustering length increases, more and more points are added to the 
peak cluster causing $\vv_p^{min}$, $\overline \vv_p$, and consequently $\nu$, to decrease. 
At $b=0.17$ the peak cluster of U70 
merges with the rest of the light curve, and the brightening of the star for HJD~$>$~2451770 
becomes undetectable, while the brightening of N188 can be detected until $b=0.3$ (but most 
probably not much farther). 

\begin{figure}[H]
   \centerline{\includegraphics[width=0.95\textwidth,
               bb = 27 295 563 687, clip]{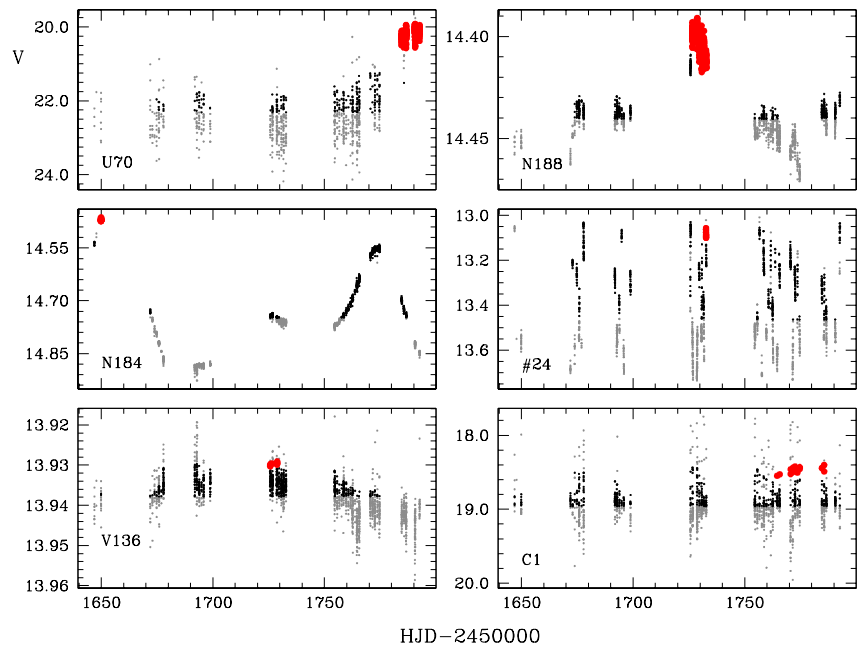}}
   \caption{Same as Fig. \ref{fig:examples1003} for $b=0.1$
   \label{fig:examples1011}}
\end{figure}

N184 is a semiregular B-type OSARG (Wray et al. 2004) whose variability was not detected by AOV 
algorithms of R17, while star \#24 exhibits strictly periodic oscillations (which are not evident 
in Figs. \ref{fig:examples1003} and \ref{fig:examples1011} because their period of 1.715~d is too 
short). In both cases (blue and cyan lines in Fig. \ref{fig:scan}) $\nu>0.75$ is observed for 
$0.1<b<0.29$ independently of $N_{\rm M}$. This somewhat surprising bonus of the method shows that 
FoF can also quite efficiently detect various cases of a more regular variability. The necessary 
condition for it to perform well on semiregular or periodic light curves is that at least one 
maximum must be well sampled.  
 
An even more surprising detection of variability occurred in the case of V136 (magenta line in Fig. 
\ref{fig:scan}). At a first glance, the light curve looks immune to FoF: Figs. \ref{fig:examples1003} 
and \ref{fig:examples1011} show a roughly sinusoidal brightness change with an amplitude barely larger 
than the observational noise, of which a half of the period is only covered. However, the densely 
populated core of the broad maximum of the curve is located above the median $\mathrm{Md}(V)$. Since 
a small $b$ probes the densest regions of the $(\tau,\vv)$ plane, for all $N_{min}$ the corresponding 
peak cluster resides in the core, yielding $\nu>0$. With a larger $b$ another, more loosely bound peak cluster
is found, located higher up in the less densely populated envelope of the maximum, causing and increase 
in~$\nu$. For $b>0.135$ the peak cluster merges with the rest of the light curve, and the subtle variability 
of V136 ceases to be detectable for FoF. The secondary maximum at $0.15<b<0.22$ and plateau at $b>0.22$ originate 
from small groups of points with largest observational errors at HJD 2451692 and HJD 2451773, respectively. 
A similar spurious variability is observed for the constant star C1 (black line). However, while for 
$N_{\rm M}=5$ it is detectable in the whole range of $b$, for increasing $N_{\rm M}$ the effect is limited to 
a narrower and narrower subrange. 

We conclude that the optimum value of $b$ lies between 0.02 and 0.04, i.e. in a range in which the 
clearly physically variable stars U70, N188, N184, and \#24 have the largest variability index.
Encouragingly, for $5\le N_{\rm M}\le 15$ the results of variability search depend rather weakly 
on the minimum size of the cluster. Some precaution is advisable, however. As Fig. \ref{fig:scan} 
indicates, adopting smaller values from this range may increase the probability of false detections, 
while at larger values the variability index of real physical variables may be slightly underestimated. 
One must also remember that even $\nu=1$ does not guarantee that the star is physically variable. 
All FoF can do is to select from a large sample of stars a smaller subsample of candidate variables 
whose light curves must be visually inspected. The question of how large this subsample should be, 
i.e. down to which $\nu$ the inspection should continue, will be answered in the next chapter.

\section{Search for new variables in the M22 field}
\label{sec:search}

Encouraged by the results of the trial runs, we decided to conduct a thorough search for new 
variables in M22. All the known variables were removed from the light curve collection of R17,
and for the remaining stars the variability index~$\nu$ was calculated using $N_{\rm M}=5$, and $b=0.03$. 
The stars were sorted according to the decreasing $\nu$, and the first 3700 light curves (i.e.  
three per cent of the total) were visually inspected. 

Twenty eight new variables and two suspected 
variables were found, whose basic parametes are given in Table~\ref{tab:fof_vars}. For the naming 
convention to agree with that of R17, they are given designations V136 (a likely member of M22),
U70 (a star whose membership status is unknown), and from N176 on (field stars). The membership
criteria are the same as in R17 - mainly based on proper motion measurements presented in Narloch 
et al. (2017; hereafter N17). Stars \#24 and C1 are included in Table \ref{tab:fof_vars} as the 
objects used for code calibration in Sect.~\ref{sec:calib}. The equatorial coordinates in columns 
2 and 3 conform to the UCAC4 system (Zacharias et al. 2013), and are accurate to 0$''$.2 -- 0$''$.3. 
Columns 4--7 contain average $V$-band magnitude $\overline V$, $\overline B-\overline V$ color, full 
amplitude in the $V$--band, and membership status. In the next two columns variablity type and 
variability index $\nu$ are given. The last column gives detectability coefficients 
$\delta=N_{\rm s}/N_{\rm lc}*100$, where $N_{\rm s}$ is the sequential number of a star on a list of stars 
sorted according to the decreasing~$\nu$, and $N_{\rm \rm lc}=123,775$ is the total number of the examined 
lighcurves (note that smaller values of $\delta$ indicate better detectability).

\begin{figure}
   \centerline{\includegraphics[width=0.95\textwidth,
               bb = 52 53 562 739, clip]{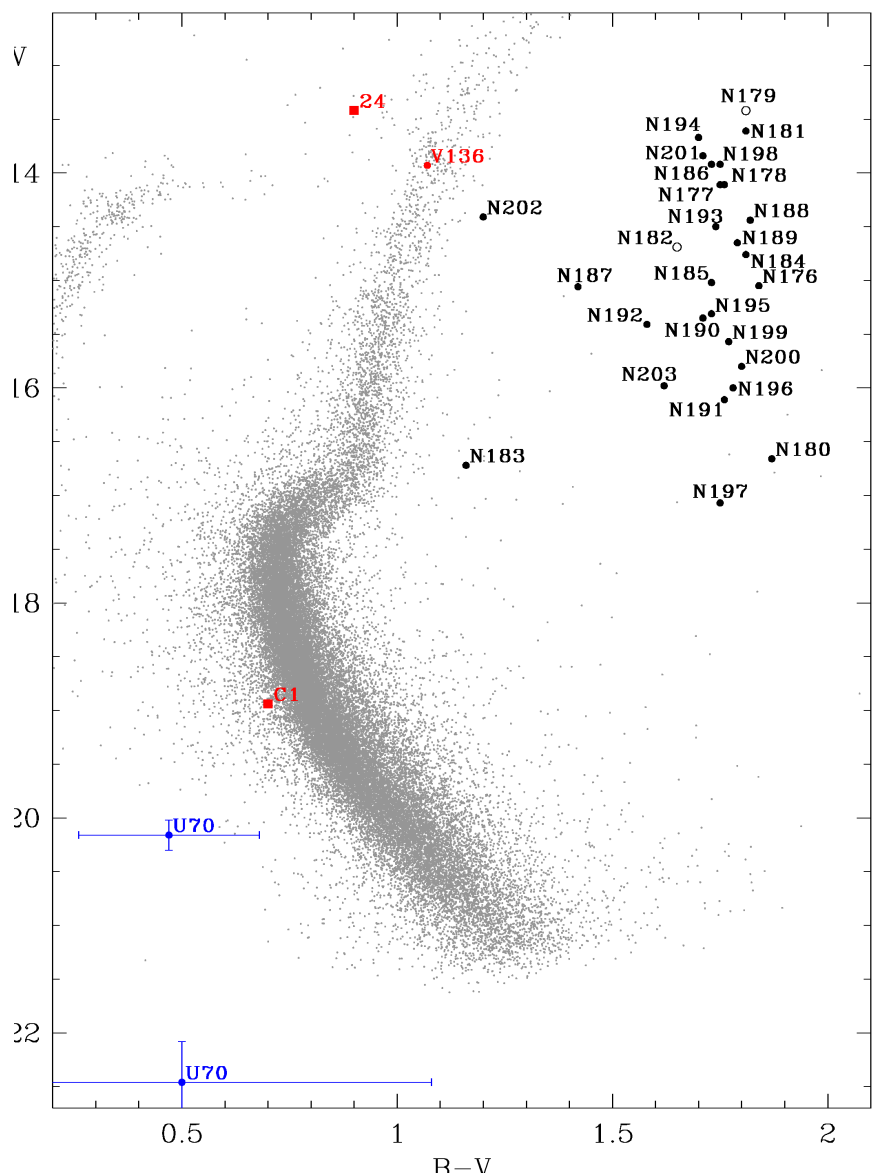}}
   \caption{CMD for the M22 field with locations of FoF-detected variables.
    Red: Members or likely members of M22; blue: stars of unknown membership; black: 
    field stars. Filled circles: confirmed variables; open circles: suspected 
    variables. Squares: stars included in Figs. \ref{fig:scan}--\ref{fig:examples1011} 
    for code-testing purposes. C1 is constant; star \#24 is a BL Her - type variable 
    described by R17. Star U70 is shown at minimum and maximum light. The gray background 
    stars are PM-members of M22 identified by N17.
    \label{fig:cmd_fof}}
\end{figure}

A CMD of M22 with locations of the new variables is shown in Fig.~\ref{fig:cmd_fof}. The gray 
background stars are PM--members of M22 identified by N17. The only new likely member of the cluster 
is the red giant V136 described in Sect. \ref{sec:calib}. The most interesting object in the sample 
is undoubtedly U70, which between HJD 2451765 and HJD 2451783 brightened by $\sim$2.3 mag, and 
remained in this state for at least 15~d. Compelling explanations for such a behavior are DN 
eruption or microlensing effect. If U70 was a member of M22, then for $(m-M)_{\rm V}=13.6$~mag and 
$E(B-V)=0.34$~mag (Harris 1996, 2010 edition) its extinction-corrected absolute magnitude at maximum 
would be $M_{\rm V}\sim$5.5 mag - a lowish, but still acceptable value for a DN (Patterson 2011). On the 
other hand, no X-ray source is located within 30\arcs\ from U70 (Rosen et al. 2016), and no indication 
for a color change between low and high state of U70 is observed, which rather speaks for the 
gravitational lens hypothesis. Unfortunately, because of the large scatter of measurements and incomplete 
coverage of the light curve we are not able to distinguish between these 
two possibilities. In any case, this is the fourth transient in the M22 field, the first three having 
been detected by Pietrukowicz et al. (2005).  

\begin{figure}[H]
   \centerline{\includegraphics[width=0.95\textwidth,
               bb = 47 423 563 687, clip]{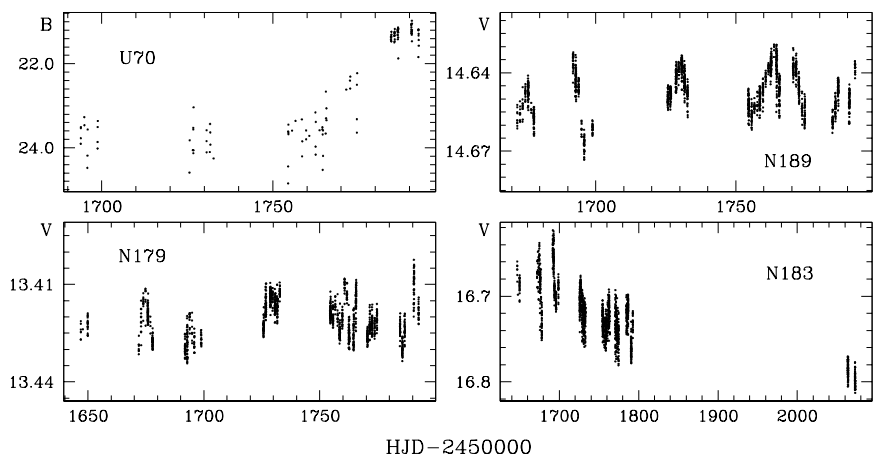}}
   \caption{Top left to bottom right: $B$-band light curve of star U70, and $V$-band 
    light curves of stars N189, N179 and N183 - examples of, respectively, OSARG-A, 
    suspected variable, and long-term variable from Table \ref{tab:fof_vars}.
    \label{fig:newvar}}
\end{figure}

Proper motions indicate that all the remaining stars from Table \ref{tab:fof_vars} except C1 and \#24 
are field objects (N17). Most of them are OSARGs of type B, evident exceptions being type 
A~OSARGs N189 and N201. The light curve of N184 in Figs. \ref{fig:examples1003} and 
\ref{fig:examples1011} is typical of OSARG-B 
objects. An example OSARG-A light curve is shown in Fig. \ref{fig:newvar} along with another two 
curves of stars from Table~\ref{tab:fof_vars}, and the $B$-band light curve of U70.\footnote{Light 
curves of all the variables from Table \ref{tab:fof_vars} are available at http://case.camk.edu.pl}

Twenty-nine variables have $\delta <1.0$, i.e. they belong to the first percentile of the whole set of 
light curves sorted according to the decreasing $\nu$. In the second percentile there is only one 
star - the suspected variable N182. Similarly, V136 is the only star in the third precentile. Based on 
these results one may expect that using FoF can reduce the number of stars that have to be individually 
inspected to about one hundredth of a given data set. An obvious practical rule is to stop inspecting 
when the function $N_{\rm var}(N_{\rm s})$ flattens, where $N_{\rm var}$ is the number of detected physical 
variables, and $N_{\rm s}$ is the sequential number of a star in the sorted data set. 

\section {Performance of FoF on OGLE-III light curves}
\label{sec:ogle}

To independently assess the potential of FoF, and estimate the efficiency with which it can detect eruptive
variables, we applied it to the set of over 345,000 light curves from twenty one OGLE-III Galactic disk 
fields (Szymanski et al. 2010), in which 40~new dwarf novae were identified by MP13. 

The whole set was divided into seven subsets differing considerably in time coverage and number of observations, 
and, to a lesser extent, in density of stellar images and/or photometric quality (see Pietrukowicz et al. 2013). 
For all stars in each subset the variability index was computed using $(N_{\rm M},\,b)$ = (5, 0.03). Next, each subset 
was sorted according to the decreasing $\nu$, sequential numbers $n_{\rm DN}$ of OGLE-III DNe in the so-obtained 
sequence were found, and detectability coefficients $\delta=n_{\rm DN}/N_{\rm lc}*100$ were calculated, where $N_{\rm lc}$ 
is the number of light curves in the subset. The results are given in the 4th column of Table~\ref{tab:OGLE1}. 
In five cases $\delta$ is larger than 1, meaning that 35 DNe (87.5\% of the total) would be found if the search 
was limited to $\delta_{max}=1\%$, i.e. to the first percentile of the whole data set. All DNe would be found 
only with $\delta_{max}\approx16\%$, implying less than a tenfold reduction in the number of the light curves to 
be inspected. 

To see how these results 
depend on control parameters, we made additional runs using $(N_{\rm M},\,b)$ = (5, 0.02) and (10, 0.02). Respective 
detectability coefficients are listed in columns 5 and 6 of Table~\ref{tab:OGLE1}. In both cases three DNe 
remain undetected, i.e. the overall detection efficiency increases to 92.5\%. Altogether, with  $\delta_{max}=1\%$ 
thirty one DNe are detected in all three $(N_{\rm M},\,b)$ runs, and further seven DNe in two runs. The remaining two DNe,
OGLE-GD-DN-011 and OGLE-GD-DN-038, are found in single runs only. Their light curves have a poorer than average 
quality, but they are not really the poorest ones in the collection of MP13. Additional factor(s) must influence
their detectability, e.g. the number of objects (including artefacts) with $\nu$ larger than that of either of 
the two DNe. 

The DN ligh tcurves of MP13 significantly differ from each other in time-coverage, average sampling density, 
and photometric accuracy, and display a rich variety of shapes. As such, they may be regarded as a fairly 
representative sample, allowing for the generalization of our findings. Thus, we expect that in surveys 
comparable in quality to OGLE about ten per cent of DNe (and, probably, of eruptive variables in general) 
may remain undetected by FoF if only one combination of control parameters $(N_{\rm M},\,b)$ is used with $\delta_{max} 
=1\%$ , i.e. if one does not intend to inspect more than 1\% of the light curves. While a detection 
efficiency of $\sim$90\% seems quite acceptable, at least in some cases it can be increased to nearly 100\% when 
a {\it combined search} involving output from runs with different control parameters is applied (note that in 
our survey each DN is detected in at least one run with $\delta_{max}=1\%$). The corresponding \texttt{Unix} 
procedure consists of the following three steps:
\begin{enumerate}
\item Sort the output from each run over decreasing $\nu$, and select the first 1\% of stars. Write $id$ and $\nu$ 
of each selected star to file A. Sort A over decreasing~$\nu$. {\it Optionally, remove stars with $\nu\approx0$ 
from A}. Write $id$s of stars from file A to file B. 
\item Apply the command \texttt{sort -u B -o B} which removes repeated $id$s from B (in our survey, B gets shorter
by up to 40\%). In the \texttt{Bash Shell} apply the command \texttt{for a in `cat B`; do grep -m1 \$a A; done > C},
which for each star from B finds the first occurrence of its $id$ in A, and writes the corresponding record to C. 
\item Sort C over decreasing~$\nu$, and overwrite B with star $id$s from C. Inspect light curves of stars from B.
\end{enumerate}
With such an arrangement, stars found by FoF to be most likely variable will be inspected first. The results of 
our combined search for DNe are displayed in Table~\ref{tab:OGLE2} which for each data subset identified in
column 1 gives the number of stars in file B (column 2), percentage of stars from file B in the data subset (column 
3), and percentage of stars from file B that would have to be inspected before {\it all} OGLE-III DNe from the given 
subset were found (column 4). As the last row of Table~\ref{tab:OGLE2} shows, all DNe can be identified at the expense 
of going through about 2\% of the light curves. One should remember, of course, that the combined search 
does not warrant a 100\% success in all cases. One can be sure, however, that in general its detection efficiency 
will be higher than that of a search based on a single combination of FoF control parameters.

\section{Summary}
 \label{sec:sum}
We have shown that the FoF algorithm adapted to the analysis of light curves can be succesfully applied 
to searches for transient or aperiodic phenomena in large photometric databases. Given the control parameters 
$N_{\rm M}$ and $b$ introduced in Section \ref{sec:impl}, our implementation of FoF assigns to each curve the 
variability parameter $\nu$ with a value between 0 and 1, with larger values indicating the presence of at 
least one brightening episode, and smaller values suggesting that the examined star does not vary in brightness. 
A test conducted on dwarf novae found by MP13 in OGLE-III data allows to estimate the detection efficiency of our 
implementation at $\sim$90\% when the search is limited to the first 1\% of light curves sorted over decreasing 
$\nu$, and only one combination of control parameters is used. A search involving $\sim$2\% of light curves and
three combinations of control parameters can be significantly more effective - in our test case all OGLE-III 
dwarf novae were identified. 

We believe FoF is a useful tool, but we are aware of its limitations. In particular, potential users must remember 
that it does not distinguish between observational artefacts and physical variability, and for this reason it may 
perform rather poorly on noisy data. Also, we cannot be sure that the values of $N_{\rm M}$ and $b$ parameters used in 
the present paper would be the best choice for all light curve collections. We rather suggest to perform a 
calibration like that in Section \ref{sec:calib} every time a new data set is analyzed.  
 
Finally, we mention that upon calibrating the control parameters we realized that FoF can also quite efficiently 
detect semi-regular or strictly periodic variability. In Section \ref{sec:search} we report 28 new variables found 
in the field of the globular cluster M22, which had been examined earlier by R17 using periodicity-searching 
algorithms. New transient, semi-regular or periodic variables found in the OGLE-III data will be reported elsewhere. 

\Acknow
{PP was supported from the Polish National Science Centre grant MAESTRO 2014/14/A/ST9/00121. We thank Joe Smak for
useful comments, and Grzegorz Pojma\'nski for the lc code which vastly facilitated the work with light curves.}

\clearpage

\begin{table}[H]
\footnotesize
 \begin{center}
 \caption{\footnotesize Basic data of FoF variables discovered in the field of M22 \strut}
          \label{tab:fof_vars}
 \begin{tabular}{|l|c|c|c|c|c|c|c|c|c|}
  \hline
 ID & RA\strut    & DEC    & $\overline V$ & $\overline B-\overline V$ & $\Delta_{\rm V}$ & Mem$^a$ & Type$^b$ & ${\nu}^c$ 
& $\delta^d$ \\
    & [deg] & [deg]  &[mag]& [mag] & [mag]      &         &        &         &  [\%]     \\
  \hline
V136  &  279.16521  &  $-23.87854$  &  13.93  &  1.07  &  0.02  &  Y  &  V  &  0.34  &  2.50  \\
U70$^e$   &  278.98646  &  $-23.90984$  &  22.46  &  0.50  &  2.30  &  U  &  V  &  0.89  &  0.04  \\
U70   &  278.98646  &  $-23.90984$  &  20.16  &  0.47  &  2.30  &  U  &  V  &  0.89  &  0.04  \\
N176  &  279.19779  &  $-23.73977$  &  15.05  &  1.84  &  0.28  &  N  &  V  &  0.93  &  0.03  \\
N177  &  279.21259  &  $-23.83521$  &  14.11  &  1.75  &  0.07  &  N  &  V  &  0.93  &  0.03  \\
N178  &  279.20565  &  $-23.91109$  &  14.11  &  1.76  &  0.13  &  N  &  V  &  0.88  &  0.04  \\
N179  &  279.19954  &  $-23.89914$  &  13.42  &  1.81  &  0.02  &  N  &  S  &  0.42  &  0.90  \\
N180  &  279.22549  &  $-23.93757$  &  16.66  &  1.87  &  0.42  &  N  &  V  &  0.93  &  0.03  \\
N181  &  279.20120  &  $-24.05648$  &  13.61  &  1.81  &  0.09  &  N  &  V  &  0.87  &  0.04  \\
N182  &  279.12222  &  $-23.80253$  &  14.69  &  1.65  &  0.03  &  N  &  S  &  0.37  &  1.53  \\
N183  &  279.13698  &  $-23.88486$  &  16.72  &  1.16  &  0.16  &  N  &  V  &  0.63  &  0.11  \\
N184  &  279.15103  &  $-23.92146$  &  14.76  &  1.81  &  0.39  &  N  &  V  &  0.98  &  0.01  \\
N185  &  279.14376  &  $-23.96482$  &  15.02  &  1.73  &  0.58  &  N  &  V  &  0.96  &  0.01  \\
N186  &  279.15241  &  $-24.01048$  &  13.92  &  1.73  &  0.12  &  N  &  V  &  0.86  &  0.04  \\
N187  &  279.11106  &  $-23.75086$  &  15.06  &  1.42  &  0.23  &  N  &  V  &  0.79  &  0.06  \\
N188  &  279.04501  &  $-23.76959$  &  14.44  &  1.82  &  0.06  &  N  &  V  &  0.97  &  0.01  \\
N189  &  279.09219  &  $-23.81836$  &  14.65  &  1.79  &  0.03  &  N  &  V  &  0.69  &  0.09  \\
N190  &  279.07769  &  $-23.83870$  &  15.35  &  1.71  &  0.21  &  N  &  V  &  0.95  &  0.02  \\
N191  &  279.07322  &  $-23.83017$  &  16.11  &  1.76  &  0.56  &  N  &  V  &  0.95  &  0.02  \\
N192  &  279.05198  &  $-23.92851$  &  15.41  &  1.58  &  0.25  &  N  &  V  &  0.88  &  0.04  \\
N193  &  279.08704  &  $-23.97307$  &  14.50  &  1.74  &  0.04  &  N  &  V  &  0.61  &  0.14  \\
N194  &  279.07205  &  $-23.98484$  &  13.67  &  1.70  &  0.15  &  N  &  V  &  0.93  &  0.03  \\
N195  &  279.07035  &  $-24.09773$  &  15.31  &  1.73  &  0.42  &  N  &  V  &  0.95  &  0.02  \\
N196  &  278.98961  &  $-23.73304$  &  16.00  &  1.78  &  0.61  &  N  &  V  &  0.97  &  0.01  \\
N197  &  279.00753  &  $-23.79984$  &  17.07  &  1.75  &  0.23  &  N  &  V  &  0.89  &  0.03  \\
N198  &  278.97410  &  $-23.84254$  &  13.92  &  1.75  &  0.31  &  N  &  V  &  0.94  &  0.02  \\
N199  &  279.01856  &  $-23.86728$  &  15.57  &  1.77  &  0.07  &  N  &  V  &  0.94  &  0.02  \\
N200  &  279.01240  &  $-23.87550$  &  15.80  &  1.80  &  0.37  &  N  &  V  &  0.90  &  0.03  \\
N201  &  279.01744  &  $-23.98125$  &  13.84  &  1.71  &  0.04  &  N  &  V  &  0.92  &  0.03  \\
N202  &  278.99578  &  $-23.99937$  &  14.41  &  1.20  &  0.04  &  N  &  V  &  0.80  &  0.06  \\
N203  &  278.97643  &  $-24.05441$  &  15.98  &  1.62  &  0.29  &  N  &  V  &  0.70  &  0.08  \\
\#24  &  279.09078  &  $-23.90371$  &  13.42  &  0.90  &  0.65  &  Y  &  V  &  0.93  &  0.03  \\
C1    &  279.08297  &  $-23.86369$  &  18.94  &  0.70  &  0.30  &  Y  &  C  &  0.13  &  77.4  \\
  \hline
 \end{tabular}
\end{center}
{\footnotesize
$^a$ Membership status: Y - member, U - no data or data ambiguous, N - field object.\\
$^b$ V - variable, S - suspected variable, C - constant star.\\
$^c$ Variability index defined in Sect. \ref{sec:impl}, obtained for $N_{\rm M}=5$ and $b=0.03$.\\
$^d$ $\delta$ is the percentage of lighcurves sorted over decreasing $\nu$ that has to be
 inspected before the star indicated in the first column is encountered.\\
$^e$ For U70, magnitude and color at low and high state are given.}
\end{table}

\clearpage

\begin{table}[H]
\footnotesize
 \begin{center}
 \caption{\footnotesize Detectability of OGLE-III galactic disk dwarf novae  \strut}
          \label{tab:OGLE1}
 \begin{tabular}{|c|l|r|r|r|r|}
  \hline
\multirow{3}{*}{Object} &\multirow{3}{*}{Data subset$^a$} &\multirow{3}{*}{$N_{\rm lc}$} & \multicolumn{3}{c|}{($N_{\rm M},\,b$)}\\
\cline{4-6}
                     & & & (5, 0.03)&(5, 0.02)&(10, 0.02) \\
                     & & & $\delta^b\;[\%]$ & $\delta\;[\%]$ & $\delta\;[\%]$  \\
  \hline
OGLE-GD-DN-001	&	CAR115-118	&	25048	&	0.03	&	0.03	&	0.01	\\
OGLE-GD-DN-002	&	CAR115-118	&	25048	&	0.55	&	0.22	&	0.06	\\
OGLE-GD-DN-003	&	CAR115-118	&	25048	&	0.01	&	0.00	&	0.02	\\
OGLE-GD-DN-004	&	CAR115-118	&	25048	&	0.05	&	0.04	&	0.04	\\
OGLE-GD-DN-005	&	CAR115-118	&	25048	&	0.38	&	0.14	&	0.04	\\
OGLE-GD-DN-006	&	CAR115-118	&	25048	&	1.01	&	0.62	&	0.17	\\
OGLE-GD-DN-007	&	CAR115-118	&	25048	&	0.45	&	0.20	&	0.06	\\
OGLE-GD-DN-008	&	CAR115-118	&	25048	&	0.28	&	0.02	&	0.10	\\
OGLE-GD-DN-009	&	CAR107-110	&	34475	&	0.12	&	0.03	&	0.01	\\
OGLE-GD-DN-010	&	CAR107-110	&	34475	&	0.01	&	0.01	&	0.03	\\
OGLE-GD-DN-011	&	CAR107-110	&	34475	&	0.20	&	2.58	&	10.09	\\
OGLE-GD-DN-012	&	CAR107-110	&	34475	&	0.22	&	0.07	&	0.06	\\
OGLE-GD-DN-013	&	CAR111-114	&	88975	&	0.15	&	0.08	&	0.04	\\
OGLE-GD-DN-014	&	CAR111-114	&	88975	&	0.15	&	0.02	&	0.00	\\
OGLE-GD-DN-015	&	CAR107-110	&	34475	&	0.06	&	0.03	&	24.17	\\
OGLE-GD-DN-016	&	CAR107-110	&	34475	&	0.35	&	0.20	&	0.05	\\
OGLE-GD-DN-017	&	CAR100-105	&	78558	&	0.30	&	0.18	&	0.05	\\
OGLE-GD-DN-018	&	CAR100-105	&	78558	&	0.94	&	0.40	&	0.11	\\
OGLE-GD-DN-019	&	CAR100-105	&	78558	&	0.09	&	0.22	&	0.07	\\
OGLE-GD-DN-020	&	CAR100-105	&	78558	&	1.28	&	0.07	&	0.19	\\
OGLE-GD-DN-021	&	CAR100-105	&	78558	&	0.90	&	0.40	&	0.21	\\
OGLE-GD-DN-022	&	CAR100-105	&	78558	&	0.23	&	0.08	&	0.02	\\
OGLE-GD-DN-023	&	CAR100-105	&	78558	&	3.40	&	0.55	&	0.14	\\
OGLE-GD-DN-024	&	CAR111-114	&	88975	&	0.25	&	0.13	&	0.05	\\
OGLE-GD-DN-025	&	CAR111-114	&	88975	&	0.12	&	0.12	&	0.04	\\
OGLE-GD-DN-026	&	CAR111-114	&	88975	&	0.53	&	0.29	&	0.08	\\
OGLE-GD-DN-027	&	CAR111-114	&	88975	&	0.04	&	0.26	&	76.19	\\
OGLE-GD-DN-028	&	CAR111-114	&	88975	&	0.04	&	0.02	&	0.06	\\
OGLE-GD-DN-029	&	CAR106		&	8710	&	0.29	&	0.24	&	0.11	\\
OGLE-GD-DN-030	&	CAR100-105	&	78558	&	0.01	&	0.00	&	0.01	\\
OGLE-GD-DN-031	&	CAR100-105	&	78558	&	0.93	&	0.55	&	0.45	\\
OGLE-GD-DN-032	&	CEN106-107	&	26271	&	0.01	&	0.01	&	0.00	\\
OGLE-GD-DN-033	&	CEN108-MUS	&	83423	&	0.36	&	0.11	&	0.04	\\
OGLE-GD-DN-034	&	CEN108-MUS	&	83423	&	0.58	&	0.04	&	0.11	\\
OGLE-GD-DN-035	&	CEN108-MUS	&	83423	&	0.67	&	0.30	&	0.11	\\
OGLE-GD-DN-036	&	CEN108-MUS	&	83423	&	1.12	&	0.49	&	0.36	\\
OGLE-GD-DN-037	&	CEN108-MUS	&	83423	&	0.02	&	0.01	&	0.20	\\
OGLE-GD-DN-038	&	CEN108-MUS	&	83423	&	15.62	&	1.01	&	0.32	\\
OGLE-GD-DN-039	&	CEN108-MUS	&	83423	&	0.11	&	1.12	&	0.47	\\
OGLE-GD-DN-040	&	CEN108-MUS	&	83423	&	0.11	&	0.02	&	0.03	\\
\hline
 \end{tabular}
\end{center}
{\footnotesize
$^a$ Identified by OGLE-III disk fields from which the light curves of MP13 were extracted.\\
$^b$ $\delta$ is the percentage of the sorted data subset indicated in the second column that has to be 
 inspected before the dwarf nova indicated in the first column is encountered. $\delta=0$ means that less
than 0.01\% of the data subset has to be inspected.}
\end{table}

\begin{table}[H]
\footnotesize
 \begin{center}
 \caption{\footnotesize Results of the combined search$^*$\strut}
          \label{tab:OGLE2}
 \begin{tabular}{|l|r|c|c|}
  \hline
\multirow{2}{*}{Data subset} & \multirow{2}{*}{$N_{\rm B}$} & $N_{\rm B}/N_{\rm lc}$ & $N_{\rm all}/N_{\rm B}$\\
                &              &     [\%]         &       [\%]        \\
  \hline
CAR100-105	&	1451	&	1.8	&	58.3	\\
CAR106		&	163	&	1.9	&	15.3	\\
CAR107-110	&	625	&	1.8	&	19.8	\\
CAR111-114	&	1932	&	2.2	&	12.1	\\
CAR115-118	&	488	&	1.9	&	58.2	\\
CEN106-107	&	575	&	2.2	&	00.5	\\
CEN108-MUS	&	1489	&	1.8	&	79.5	\\
Total           &       6723    &       1.9     &               \\
  \hline
 \end{tabular}
\end{center}
{\footnotesize
$^*$ See text for explantions.}
\end{table}

\end{document}